# On the wave propagation analysis and supratransmission prediction of a metastable modular metastructure for non-reciprocal energy transmission

Z. Wu[*], and K.W. Wang

*Department of Mechanical Engineering, University of Michigan, Ann Arbor, MI 48109-2125*

*\* Corresponding author, email: wuzhen@umich.edu*

**Abstra**

In this research, we investigate in-depth the nonlinear energy transmission phenomenon in a metastable modular metastructure and develop efficient tools for the design of such systems. Previous studies on a one-dimensional (1D) reconfigurable metastable modular chain uncover that when the driving frequency is within the stopband of the periodic structure, there exists a threshold input amplitude, beyond which sudden increase in the energy transmission can be observed. This onset of transmission is caused by nonlinear instability and is known as supratransmission. Due to spatial asymmetry of strategically configured constituents, such transmission thresholds could shift considerably when the structure is excited from different ends and therefore enabling the non-reciprocal energy transmission. This one-way propagation characteristic can be adaptable via reconfiguring the metastable modular system. In this new study, we build upon these findings and advance the state of the art by (a) exploring the different mechanisms that are able to activate the onset of supratransmission and their implications on wave energy transmission potential, and (b) developing an effective design tool - a localized nonlinear-linear model combined with harmonic balancing and transfer matrix analyses to analytically and efficiently predict the critical threshold amplitude of the metastable modular chain. These investigations provide important new understandings of the rich and intricate dynamics achievable by nonlinearity, asymmetry, and metastability, and create opportunities to accomplish adaptable non-reciprocal wave energy transmission.

## 1   Introduction

Non-reciprocity of wave propagation refers to the unidirectional wave transmission between two points in space and it necessitates breaking the time-reversal symmetry of the system [1] [2]. Motivated by one-way flow of electrical energy using diodes, in recent years, significant research attention have been devoted to realize non-reciprocal wave propagation in other energy forms such as acoustic [3] [4] [5] [6] [7] [8] [9] [10] [11], elastic [12] [13], thermal [14] [15] and optical energy [16] [17]. These contributions can be loosely categorized into two domains: using linear systems with symmetry breaking mechanism [5] [10] [12] [17] or utilizing nonlinear systems [3] [4] [6] [7] [8] [9] [11] [13] [14] [15] [16]. While interesting, many of these studies only focused on designing systems that exhibit unidirectional energy transmission and limited attention has been paid to systems that are capable of attaining *on-demand tuning* of such non-reciprocal wave propagations, which are beneficial in many engineering applications [6] [18]. Boechler et al demonstrated tunable rectification factor by adjusting the static load of a defected granular chain [3]. Chen and Wu presented a tunable topological insulator using a 2D phononic crystal through both analysis and numerical experiments [19]. Recently, Raney et al also realized a tunable soft mechanical diode capable of one-way solitary wave propagation using elastomeric bistable lattices [20]. Amongst these studies [3] [19] [20], the tunable non-reciprocal wave propagation was achieved by adjusting global topology or external constraint of the proposed periodic structure. On the other hand, individual reconfiguration of the numerous repeating internal modules/components for a fixed global confinement, which can significantly expand the adaptation space and greatly benefit adaptive wave control, are not allowed with these approaches.

Recently, we proposed a novel approach to accomplish adaptable non-reciprocal wave propagation for a broad frequency range with large adaptation space [21]. This was achieved by exploiting the nonlinear *supratransmission* property of a spatially asymmetric reconfigurable metastable modular metastructure, systems that exhibit coexisting metastable states for the same topology [21]. Supratransmission refers to a wave propagation phenomenon that is unique to nonlinear periodic structures. For spatial periodic linear systems, there exhibits band-pass characteristics of energy transmission owning to the existence of bandgaps in the frequency spectrum. When excitation frequency is outside the bandgaps, vibrational energy is able to transmit through the structure with propagation speed determined by the input frequency. Whereas when frequency is inside the bandgaps, structure will scatter or absorb the energy and prohibit the transmission in all or specific directions [22] [23]. Such characteristic has been widely used in the development of mechanical filters [24] [25]. Nonlinear supratransmission, on the other hand, describes a sudden transmission of wave energy in periodic nonlinear structure subjected to continuous boundary driving with input frequency inside the linear bandgaps [26]. For small excitation amplitude, similar to a linear system, injected energy is spatially attenuated away from the input and does not propagate through the chain. When input exceeds a threshold amplitude, even though the excitation frequency is still inside the linear bandgaps, energy transmission becomes possible due to nonlinear instability [27]. This intriguing phenomenon has been observed and extensively investigated in many discrete nonlinear systems, such as sine-Gordon and Klein-Gordon chains of coupled oscillators [26] [28] [29], Fermi-Pasta-Ulam (FPU) nonlinear chains [30], and periodic bistable chains [31] [32]. Due to the symmetric nature of the systems in these studies, non-reciprocal wave energy transmission is not possible. To facilitate unidirectional energy transmission, we proposed a metastable module as the building block [21]. The unit module which consists of a bistable spring connected in series with a linear spring, hence the spatial symmetry is broken on both the module level and system level [33] [34]. We provided experimental and numerical evidences on non-reciprocal wave propagation using a metastable chain realized by connecting the modules in series and we discovered that due to spatial asymmetry, this non-reciprocal energy propagation is facilitated by triggering the onset of nonlinear supratransmission at different input amplitude levels when metastructure is excited in opposite directions [21]. In addition to non-reciprocal wave propagation, endowed with metastability, we discovered that the proposed metastructure can exhibit on-demand tuning and adaptation of non-reciprocal wave propagation characteristics by switching amongst metastable states for a large parameter space. Our prior investigation [21] presented a novel approach to achieve adaptable non-reciprocal wave propagation and showed great potential, however deep understanding of the proposed system is still lacking and an efficient tool is needed to design complex systems consists of the proposed modules for non-reciprocal wave energy propagation and adaptation.

To advance the state of the art, building upon the authors' foundational idea, the objectives of this research are two-folds. First, through in-depth numerical studies, we seek to explore the different mechanisms that are able to activate the onset of supratransmission and their implications on wave energy transmission potential. Second, we aim to develop effective analytical tools to analyze and synthesize systems capable of attaining non-reciprocal wave propagation for desired operation ranges. In doing so, this investigation presents an opportunity to thoroughly explore the rich dynamics of the proposed system and provides design guidelines for creating real-world engineering systems with desired non-reciprocal characteristics.

To present the approach and outcomes, this paper is organized as follows. In Sec II, the equations describing the nonlinear dynamics of the metastructure are presented, followed by dispersion analysis of the linearized system. In Sec. III, we present in detail different route to supratransmission the proposed metastructure imparts. In Sec IV, we discuss a localized nonlinear-linear model that can analytically predict the critical threshold amplitude for supratransmission and exemplify non-reciprocal characteristics of the proposed system. Lastly in Sec. V, influence of some key parameters on the threshold amplitude to trigger supratransmission is explored.

## 2 Mathematical model and governing equations

Figure 1 depicts a schematic of discrete lattice representation of *N* identical metastable modules connected in series. Each metastable module, highlighted with red dashed box, consists of two masses $m_1$ and $m_2$ (represented with orange and yellow dots) coupled via a linear constituents represented with coil springs, highlighted with blue dashed box; the modules are inter-connected by bistable springs, exemplified with buckled beams, green dashed box. Free length configuration of the structure $L_{free}$ is defined to be the zero force position when all the bistable elements are buckled to the left and the global displacement z is defined as the additional deformation of the structure starting from the free length $L_{free}$, Figure 1(a). Without loss of generality, the bistable and linear restoring forces are assumed of the form, $F_{NL} = k_1 x + k_2 x^2 + k_3 x^3$ and $F_L = k_L y$, where $x$ and $y$ are the deformations of bistable and linear springs respectively. The total potential energy of the metastable chain for a fixed global displacement $z$ measured from its free length $L_{free}$ can be expressed as:

$$U = \sum_{i=1}^{N-1}\left[\frac{k_L}{2}\left(x_{[i]2} - x_{[i]1}\right)^2 + \frac{k_1}{2}\left(x_{[i+1]1} - x_{[i]2}\right)^2 + \frac{k_2}{3}\left(x_{[i+1]1} - x_{[i]2}\right)^3 + \frac{k_3}{4}\left(x_{[i+1]1} - x_{[i]2}\right)^4\right] +$$
$$\frac{k_L}{2}\left(z - x_{[N]1}\right)^2 + \frac{k_1}{2}x_{[1]1}^2 + \frac{k_2}{3}x_{[1]1}^3 + \frac{k_3}{4}x_{[1]1}^4 \quad (1)$$

which is a function of internal mass displacements $x_{[i]1}, x_{[i]2}$, where subscript $i$ in square bracket refers to the $i^{th}$ metastable module. All internal displacements $x_{[i]1}$ and $x_{[i]2}$ are measured from the individual positions of the free length configuration. For a fixed global displacement z, the equilibrium positions of metastructure satisfies $\partial U/\partial x_{[i]1} = 0$ and $\partial U/\partial x_{[i]2} = 0$ under the constraint that $\sum_{i=1}^{N-1}(x_{[i]1} + x_{[i]2}) + x_{[N]1} = z$. Based on the minimum potential energy principle [35] [36], metastable states of the chain satisfy $\partial^2 U/\partial x_{[i]k}\partial x_{[j]l} > 0$, i.e. the Hessian matrix of the potential is positive definite.

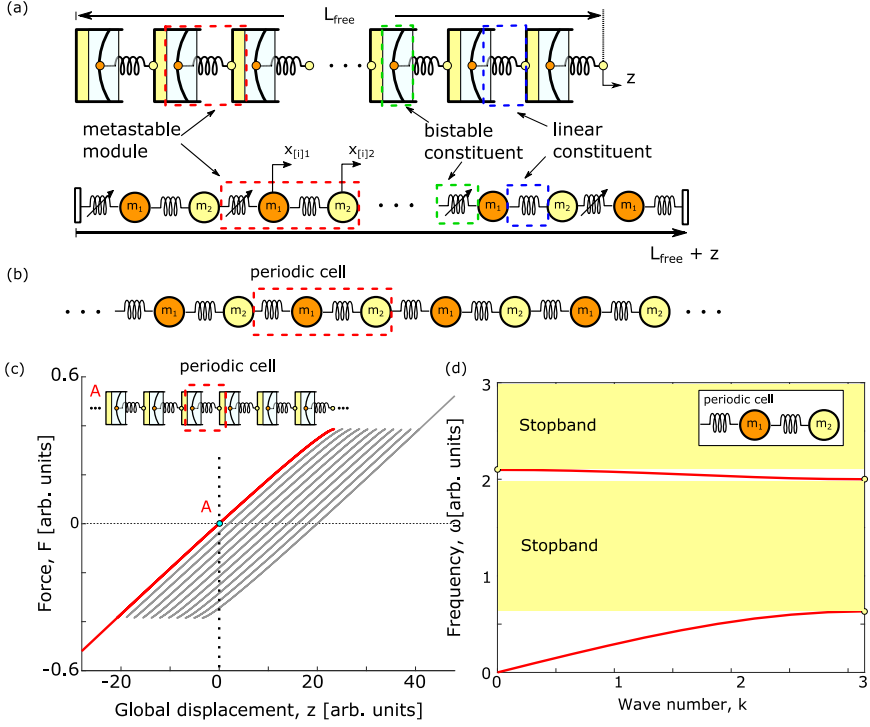

Figure 1. (a) Schematic and discrete mass spring representation of an $N$ metastable module assembled in series. Metastable module, bistable and linear constituents are highlighted with dashed boxes. (c) Force displacement profile of a 10 module metastructure assembly. (b) and (d) Linearized periodic structure for dispersion analysis and corresponding band structure of the metastructure with free length $z = 0$ for configuration A in (c).

In general, for a fixed global displacement $z$, a chain of $N$ metastable modules can have up to $2^N$ metastable states (internal configurations), Figure 1(c) [33] [34]. Starting from one of the metastable states, equations of motion for the $i^{th}$ metastable module can be expressed as:

$$m_1 \ddot{x}_{[i]1} + F_{NL}(x_{[i]1} - x_{[i-1]2}) + c(\dot{x}_{[i]1} - \dot{x}_{[i-1]2}) + k_L(x_{[i]1} - x_{[i]2}) + \zeta(\dot{x}_{[i]1} - \dot{x}_{[i]2}) = 0 \qquad (2a)$$

$$m_2 \ddot{x}_{[i]2} + F_{NL}(x_{[i]2} - x_{[i+1]1}) + c(\dot{x}_{[i]2} - \dot{x}_{[i+1]1}) + k_L(x_{[i]2} - x_{[i]1}) + \zeta(\dot{x}_{[i]2} - \dot{x}_{[i]1}) = 0 \qquad (2b)$$

where $c$ represents the damping coefficients. Eq. (2a) is applicable to $\forall i = 2$ to $N - 1$ and Eq. (2b) is applicable to $\forall i = 1$ to $N - 1$.

To investigate the non-reciprocal effect, two actuation scenarios are considered: one is forward actuation with input $x_{in}$ on the left side of the lattice chain and the other is backward actuation with input $x_{in}$ on the right hand side of the chain. Discrete representations of the excitation scenarios are depicted in Figure 2. For both scenarios, displacement input $x_{in}$ is directly applied to the mass next to the boundary of the chain and output signal $x_{out}$ is measured one unit (two degrees of freedom) away from the other side of the boundary. Owning to asymmetry of the metastable module, periodic repeating unit for the two excitation scenarios are inherently different, highlighted with red dashed boxes respectively, Figure 2. With fixed boundary conditions and displacement input $x_{in}$, for forward actuation, equations of motion of the last mass in the chain can be modified as

$$m_1\ddot{x}_{[N]1} + F_{NL}(x_{[N]1} - x_{[N-1]2}) + c(\dot{x}_{[N]1} - \dot{x}_{[N-1]2}) + k_L(x_{[N]1} - z) + \zeta(\dot{x}_{[N]1} - \dot{z}) = 0 \quad (3a)$$

and displacement of the first mass is prescribed by $x_{in}$. Similarly, for backward actuation, equations of motion for first mass can be expressed as:

$$m_1\ddot{x}_{[1]1} + F_{NL}(x_{[1]1}) + c(\dot{x}_{[1]1}) + k_L(x_{[1]1} - x_{[1]2}) + \zeta(\dot{x}_{[1]1} - \dot{x}_{[1]2}) = 0 \quad (3b)$$

To determine the dispersion relation, we first linearize the equations of motion about its metastable state with damping coefficient $c = 0$ and the band structure is determined by modeling a repeating periodic unit cell of an unforced, infinite chain, Figure 1(b). For the diatomic chain depicted in Figure 1(a), the periodic unit cell is the same as a linearized metastable module, Figure 1(b), and the linearized equation can be written as:

$$m_1\ddot{\zeta}_i + \tilde{k}_{NL}(\zeta_i - \eta_{i-1}) + k_L(\zeta_i - \eta_i) = 0 \quad (4a)$$

$$m_2\ddot{\eta}_i + \tilde{k}_{NL}(\eta_i - \zeta_{i+1}) + k_L(\eta_i - \zeta_i) = 0 \quad (4b)$$

where $\zeta_i$ and $\eta_i$ are small perturbation of mass $m_1$ and $m_2$ around its initial positions and $\tilde{k}_{NL}$ is the corresponding linearized stiffness of the bistable spring depending on the starting metastable states. Assuming solutions in the form of a traveling wave, i.e, $\zeta_i = A exp[j(\omega t - kiL)]$ and $\eta_i = B exp[j(\omega t - k(i+1)L)]$, where $k$ is the wave number and L is unit length, the model is reduced to a standard eigenvalue problem:

$$\begin{bmatrix} (k_L + \tilde{k}_{NL})/m_1 & -(\tilde{k}_{NL} + k_L e^{-jkL})/m_1 \\ -(\tilde{k}_{NL} + k_L e^{jkL})/m_2 & (k_L + \tilde{k}_{NL})/m_2 \end{bmatrix} \begin{bmatrix} A \\ B \end{bmatrix} = \omega^2 \begin{bmatrix} A \\ B \end{bmatrix} \quad (5)$$

The band structure can then be determined by sweeping the wave number $k$ from $0/L$ to $\pi/L$. It can be determined that the bandgaps are within $[\sqrt{\omega_1}, \sqrt{\omega_2}]$ and $[\sqrt{\omega_3}, \infty)$, where $\omega_1 = \frac{(k_L+\tilde{k}_{NL})(m_1+m_2)}{2m_1m_2} - \frac{\sqrt{(k_L+\tilde{k}_{NL})^2(m_1+m_2)^2 - 16k_L\tilde{k}_{NL}m_1m_2}}{2m_1m_2}$, $\omega_2 = \frac{(k_L+\tilde{k}_{NL})(m_1+m_2)}{2m_1m_2} + \frac{\sqrt{(k_L+\tilde{k}_{NL})^2(m_1+m_2)^2 - 16k_L\tilde{k}_{NL}m_1m_2}}{2m_1m_2}$, and $\omega_3 = \frac{(k_L+\tilde{k}_{NL})(m_1+m_2)}{m_1m_2}$. For demonstration purposes and without loss of generality, parameters used in the analysis are chosen to be of arbitrary unit. With $k_1 = 2, k_2 = -3, k_3 = 1, k_L = 0.2, m_1 = 1$ and $m_2 = 1$, band structures of the diatomic metastable chain in free length and zero external force configuration depicted by point A is shown in Figure 1(c). Two bandgaps of the diatomic chain are [0.633, 2] and [2.098, ∞).

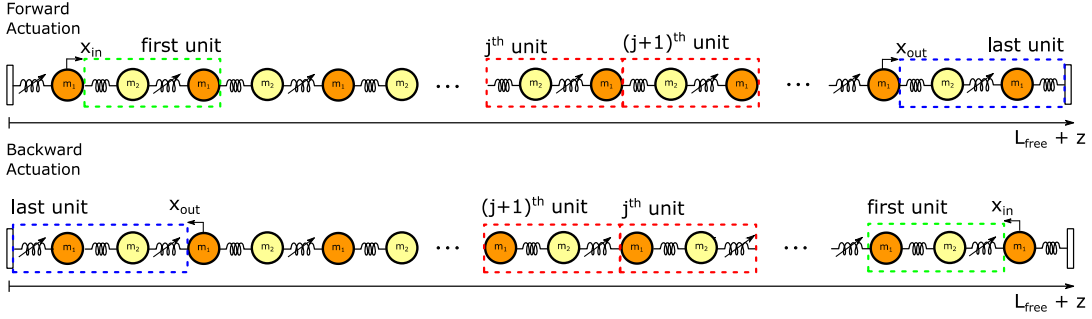

Figure 2. Conceptual diagrams of metastable module assembled in series under forward (excitation from left) and backward (excitation from right) actuations. For both scenarios, displacement input $x_{in}$ is directly applied to the mass next to the boundary of the chain, and output signal $x_{out}$ is measured one module away from the boundary. Repeating unit for both excitation scenarios are highlighted with red dashed box. The first and last units are highlighted with green and blue dashed boxes, respectively.

## 3 In-depth understanding of route to supratransmission of the metastructure – A numerical study

With the same system parameters used to determine the bandgaps in Sec. 2.1, $k_1 = 2, k_2 = -3, k_3 = 1, k_L = 0.2$ and masses $m_1 = m_2 = 1$, wave propagation characteristics are explored using a 10 module metastructure over a wide spectrum of input parameters starting from free length configuration, i.e. $z = 0$. Given the system parameters, distance between the two equilibrium positions for the bistable constituents is 2. Small damping coefficients $c = 0.001$ is applied between lattices and input $x_{in} = p\sin(\omega t)$ is prescribed to one end of the structure depending on excitation scenario with numerical simulations running for sufficiently long time (30000 periods where one period is defined as $2\pi/\sqrt{k_1/m_1}$) to reach steady state. Small damping coefficients are chosen to limit the influence of energy dissipation on wave propagation phenomenon. Figure 3 depicts the contour plot of the transmittance ratio (TR) of the metastructure as input frequency and amplitude varies for forward actuation scenario, Figure 2. The transmittance ratio (TR) is defined as the ratio of output RMS displacement over input RMS displacement $TR = |RMS(x_{out})|/|RMS(x_{in})|$. Hence, a higher transmittance ratio indicates more energy is transmitted through the metastructure. Contour plot is in log scale with brighter region corresponds to larger TR values. As shown in Figure 3, for input frequency inside the passband (PB) of the structure, since wave energy is able to propagate through the chain [23], transmittance ratio (TR) is always large. When excitation frequency is inside the stopband of the structure, an amplitude dependent wave transmission characteristic can be observed due to supratransmission [21] [26] [30] [37]. Upon close inspection, three distinct regions, labeled with I, II and III in Figure 3, with different input amplitude dependency can be identified. In this section, we will explore in detail different mechanisms that trigger the supratransmission phenomenon, which enable the energy transmission inside the bandgap.

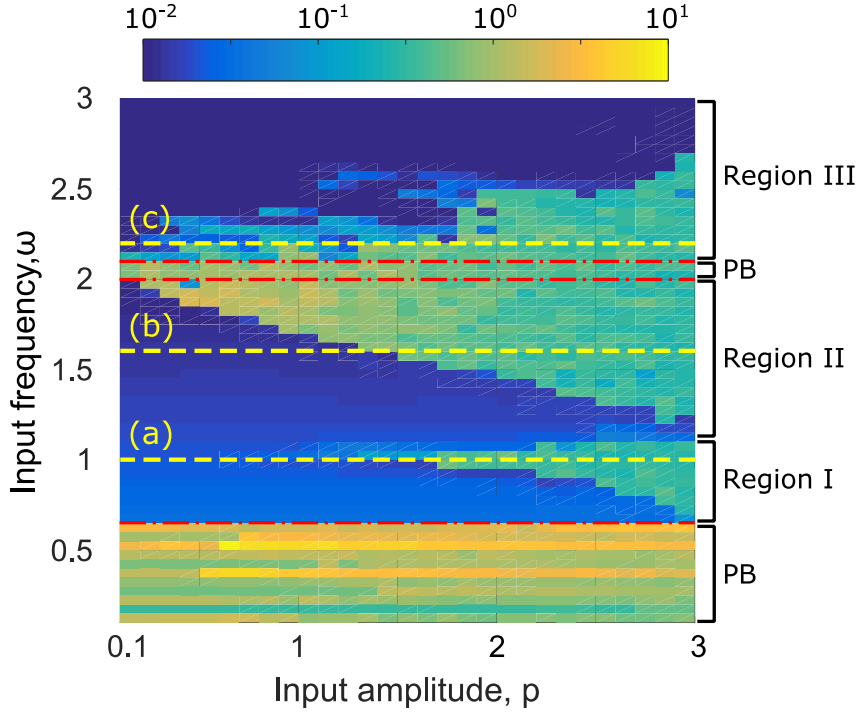

Figure 3. Contour plot on transmittance ratio (vs. input frequency and amplitude) for forward actuation from free length configuration. Passbands (PB) from dispersion analysis are within $(0, 0.632)$ and $(2, 2.098)$. Passbands are bounded by the red dash-dotted lines. For frequency inside the bandgaps, three regions, each with unique energy transmission characteristics depending on input frequency and amplitude can be identified, labeled with I, II and III. Representative frequency for each region is labeled with (a), (b) and (c) with yellow dashed lines, respectively.

*3.1 Region I*

To explore the characteristic dynamics in region I, input frequency $\omega = 1$ labeled with (a) in Figure 3 is chosen while excitation amplitude varies from 0.1 to 3 for forward actuation, Figure 2. From previous study [21], we find that energy transmission is closely related to the dynamics of the first unit near the excitation; therefore, Figure 4 depicts both the displacement amplitudes $Y$ of the last 100 excitation periods of the first bistable constituent and corresponding transmittance ratio as input amplitude increases. The *displacement amplitude* is defined to be the peak value of relative displacement between $m_1$ and $m_2$ at steady state and is chosen to represent the dynamics of the first metastable unit. If the steady state response of the bistable constituent is periodic oscillation, displacement amplitude is represented with a single point in Figure 4 for a fixed excitation level, whereas for aperiodic or chaotic oscillation, amplitude at steady state covers a wide range of magnitude over the last 100 excitation periods and therefore is represented by a line. As input amplitude varies, 6 input levels $p = [0.15, 0.45, 0.8, 1.5, 1.77, 1.87]$, labeled A through F, are carefully inspected. Figure 5 depicts the phase diagram of the bistable displacement $y$ and corresponding bistable velocity $v$ of the first unit as well as the FFT of bistable velocity $v$. Bistable displacement $y$ is defined to be the relative displacement between two masses at steady state and likewise bistable velocity $v$ is the relative velocity. Black dots represent the equilibrium positions of the bistable constituents and red dots correspond to the stroboscopic map of the first bistable constituent.

With small excitation level $p = 0.15$, point A in Figure 4, the first bistable constituent undergoes intrawell oscillation (small amplitude oscillation around one of the equilibrium positions) with dominant frequency the same as the input frequency, Figure 5(A). Since input frequency is inside the stopband, transmittance ratio is small as expected, Figure 4(b), indicating that energy does not propagate through the chain. As excitation increases to $p = 0.45$, point B in Figure 4, displacement amplitude bifurcates into two magnitudes. FFT of bistable velocity in Figure 5(B) reveals that the vibration exhibits characteristics of superharmonic oscillation with dominant frequency twice as much as the driving frequency. Since all frequency spectrums with dominant magnitude are in the stopband, the TR remains to be low. Further increasing excitation level, steady state magnitude of the first bistable element becomes aperiodic. For instance, at $p = 0.8$, intrawell aperiodic oscillation occurs with frequency contents covering a broader spectrum and some of the frequency content now are inside the passband, Figure 5(C) and hence transmittance ratio starts to increase to 0.04, indicating that vibration energy starts to propagate through the chain. As input level increases to approximately 1.3, displacement amplitude suddenly jumps and the vibration becomes chaotic interwell oscillation (large amplitude oscillation encompassing both equilibrium positions), point D and F in Figure 4. FFT of corresponding bistable velocity in Figure 5(D) and (F) also illustrates that vibration energy of the constituent is diffused to a broader frequency spectrum with more frequency content inside the passband. Hence, such interwell chaotic oscillation leads to a noticeable increase in the transmittance ratio to 0.617 and 0.845 respectively for D and F, Figure 4(b). Lastly, at excitation $p = 1.77$, displacement amplitude encompass a shorter range near one of its equilibrium positions at $y = 2$ (the other equilibrium position is at $y = 0$), point E in Figure 4(a), indicating it undergoes intrawell oscillation. Phase diagram in Figure 5(E) also corroborates this observation. From the FFT of bistable velocity, Figure 5(E), it is determined that bistable constituent of the first unit undergoes a quasi-periodic intrawell oscillation, with more energy localized on frequency contents in the stopband compare to chaotic interwell oscillations, point D and F in Figure 4. Therefore, transmittance ratio at this excitation level drops correspondingly, point E in Figure 4(b). These observations illustrate that the transmittance ratio is closely related to the dynamics of first unit. It is generally high when the first bistable constituent undergoes interwell or intrawell aperiodic/chaotic oscillation. Furthermore, for frequency range inside region I, supratransmission is shown to be facilitated through the nonlinear instability by transitioning from superharmonic vibration to aperiodic intrawell oscillation of the bistable constituent and transmittance ratio grows progressively as input amplitude increases.

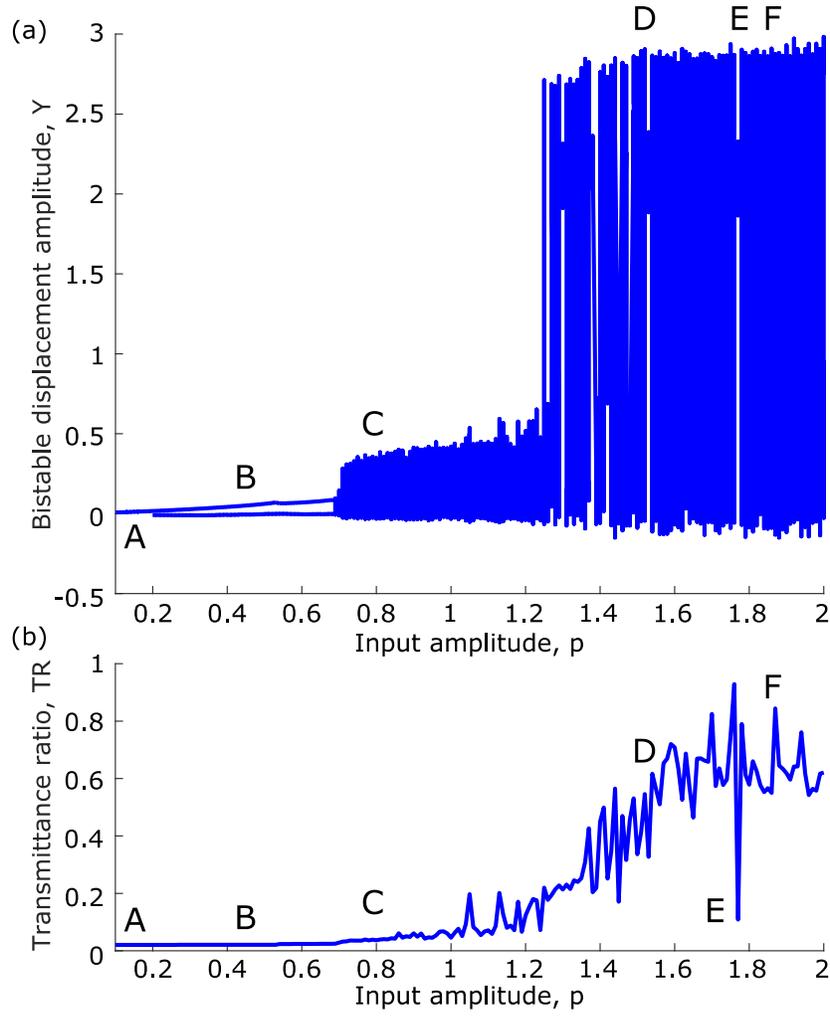

Figure 4. Displacement amplitude of the first bistable constituent and corresponding transmittance ratio as input amplitude increases for forward excitation scenarios with input frequency $\omega = 1$. Displacement amplitude is determined to be the peak value of relative displacement between the two oscillators at steady state. A through F corresponds to input amplitude $p = [0.15, 0.45, 0.8, 1.5, 1.77, 1.87]$.

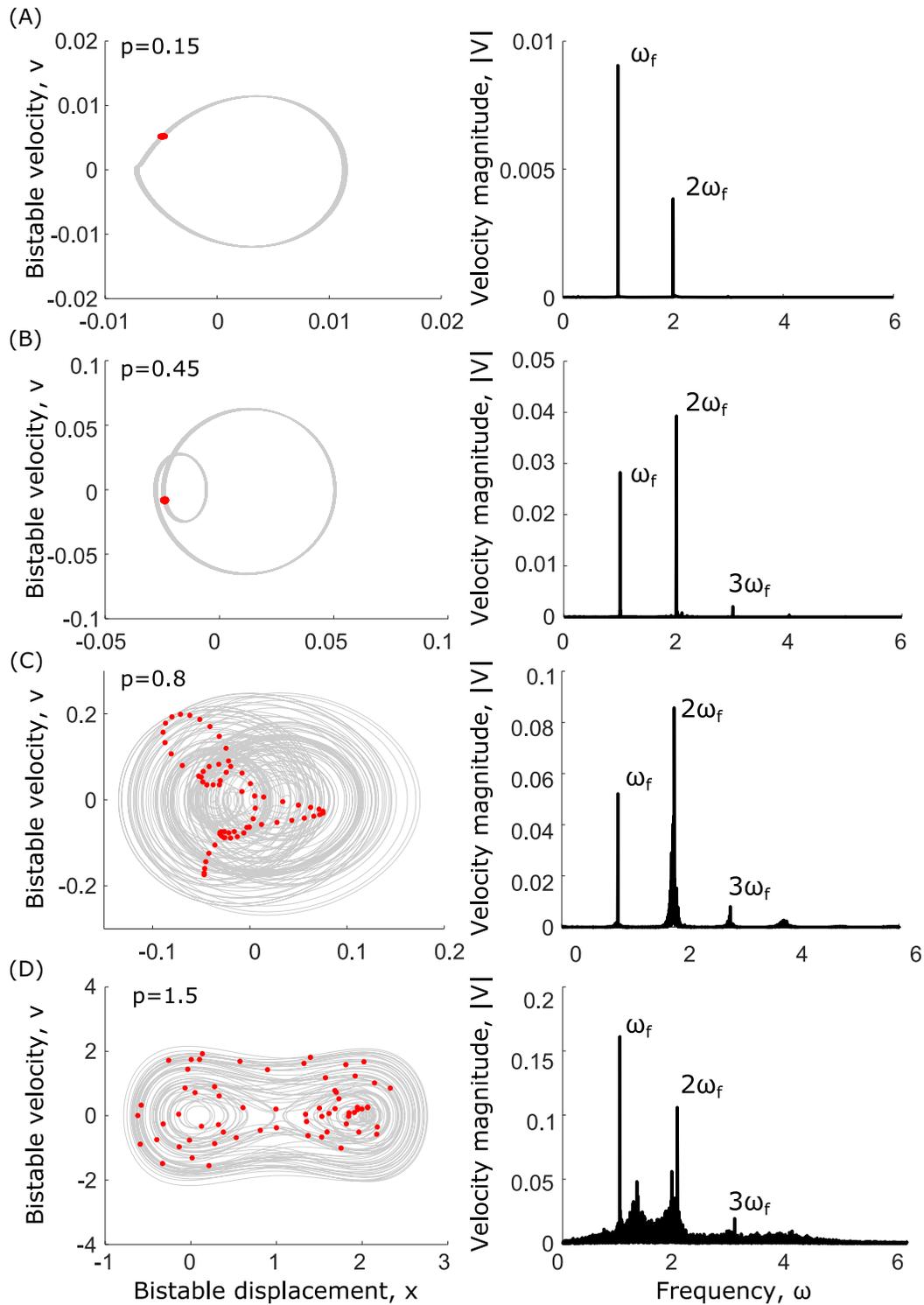

Figure 5. Phase diagram of the bistable constituent of the first unit under forward excitation, Figure 2, and corresponding FFT of the bistable velocity for different input excitation levels. (A) through (F) matches the same labels in Figure 4. Black dots represent the equilibrium positions of the bistable constituents and red dots correspond to the stroboscopic map of the first bistable constituent. $\omega = 1$ denotes the input frequency.

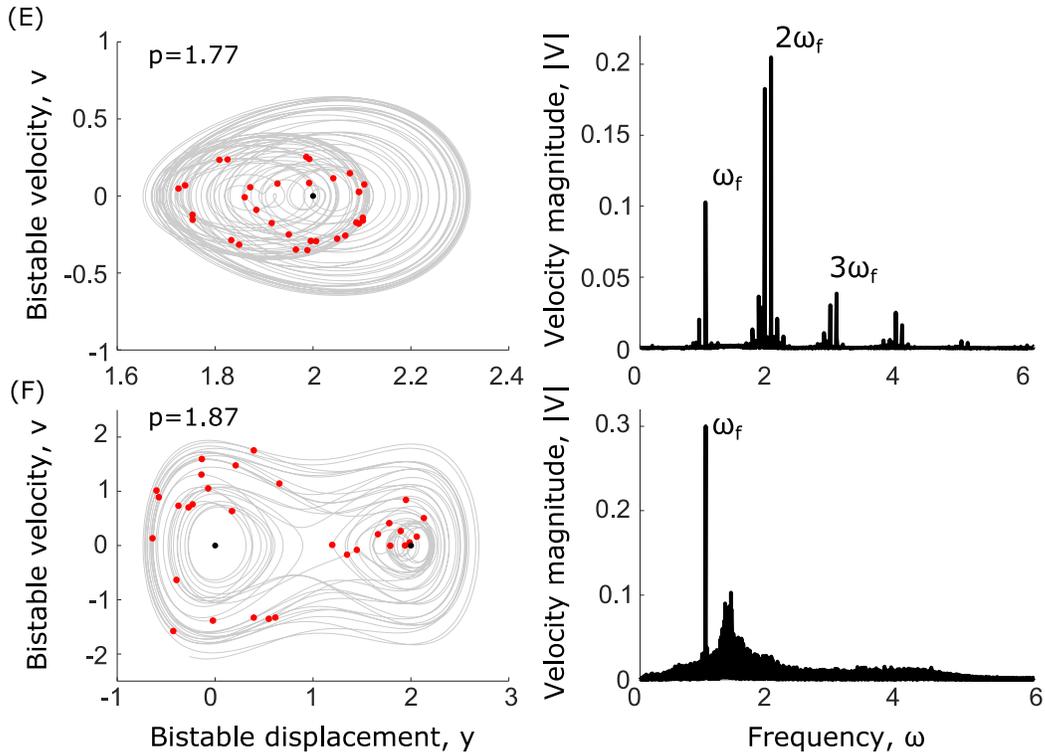

Figure 5. (Cont'd) Phase diagram of the bistable constituents of the first unit under forward excitation and corresponding FFT of the bistable velocity for different input excitation levels.

*3.2 Region II*

Similarly, to explore the characteristic dynamics in region II, input frequency $\omega = 1.6$ labeled with (b) in Figure 3 is chosen while excitation amplitude varies from 0.1 to 3 for forward actuation, Figure 2. Figure 6 depicts the displacement amplitude of the first bistable unit and corresponding transmittance ratio as input amplitude increases. As input amplitude varies, representative input levels $p = [0.5, 1.27, 1.46, 1.85]$, are selected and labeled A through D. Figure 7 depicts the phase diagram of the bistable constituent for the first unit and FFT of corresponding bistable velocity. Red dots correspond to the stroboscopic map of the first bistable constituent.

Comparable to cases in region A, when excitation level is small, transmittance ratio is low, for instance $p = 0.5$ point A in Figure 6. As shown in Figure 6(A), the first bistable element undergoes a small amplitude periodic intrawell oscillation with dominant frequency the same as the input frequency and corresponding transmittance ratio is only 0.01. As input amplitude increases to $p = 1.27$, vibration amplitude of the first bistable constituent increases significantly and the resultant vibration is characterized as chaotic interwell oscillation, Figure 7(B) and (D). Correspondingly, more than two orders of magnitude increases in transmittance ratio is observed as dynamical characteristics of the first bistable constituent suddenly changes. Transitioning from intrawell to interwell oscillation for a bistable element as input amplitude increases is known to be caused by a saddle node bifurcation [38] [39] [40], therefore, for frequency range inside region II, supratransmission is triggered through the nonlinear instability of a saddle node bifurcation. Note that point C in Figure 6 corresponds to a near periodic large amplitude interwell oscillation yet the transmittance ratio is lower compared to the chaotic interwell oscillation case point B or D. This is because when first bistable element undergoes chaotic oscillation, it converts the frequency spectrum from the stopband to the passband more effectively, leading to more energy propagation through the chain and hence resulting in a higher TR [21] [32].

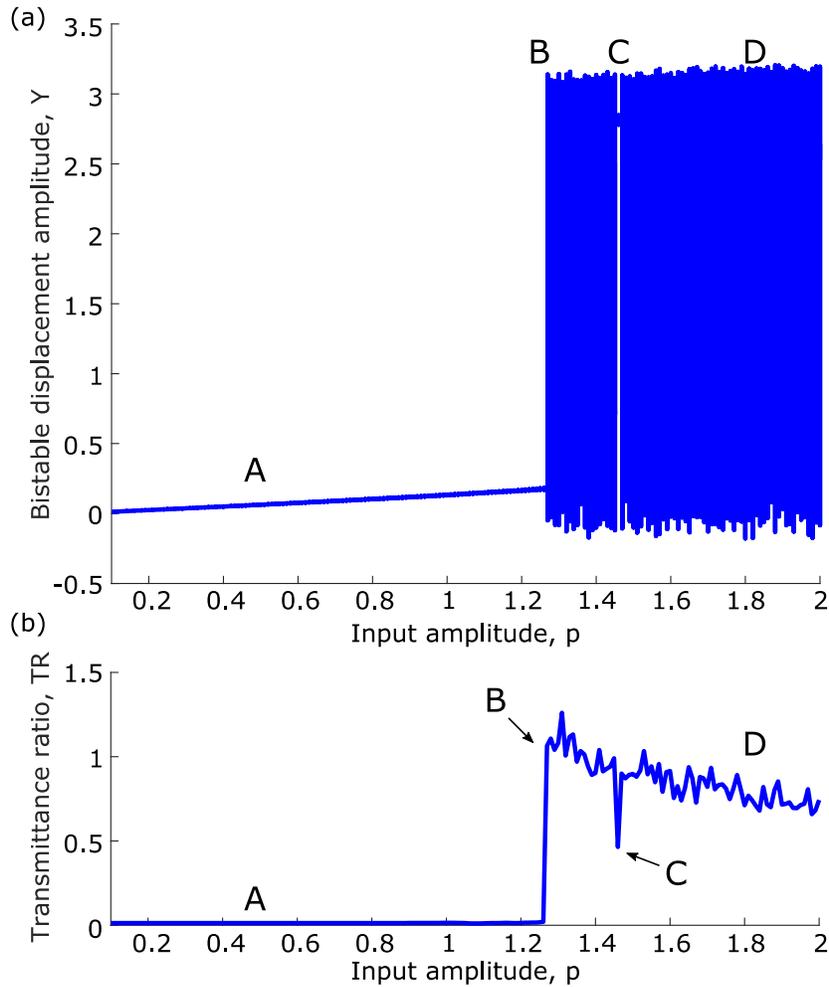

Figure 6. Displacement amplitude of the first bistable constituent and corresponding transmittance ratio as input amplitude increases for forward excitation scenarios with input frequency $\omega = 1.6$. Displacement amplitude is determined to be the relative displacement amplitude between the two moving masses at steady state. A through D corresponds to input amplitude $p = [0.5, 1.27, 1.46, 1.85]$.

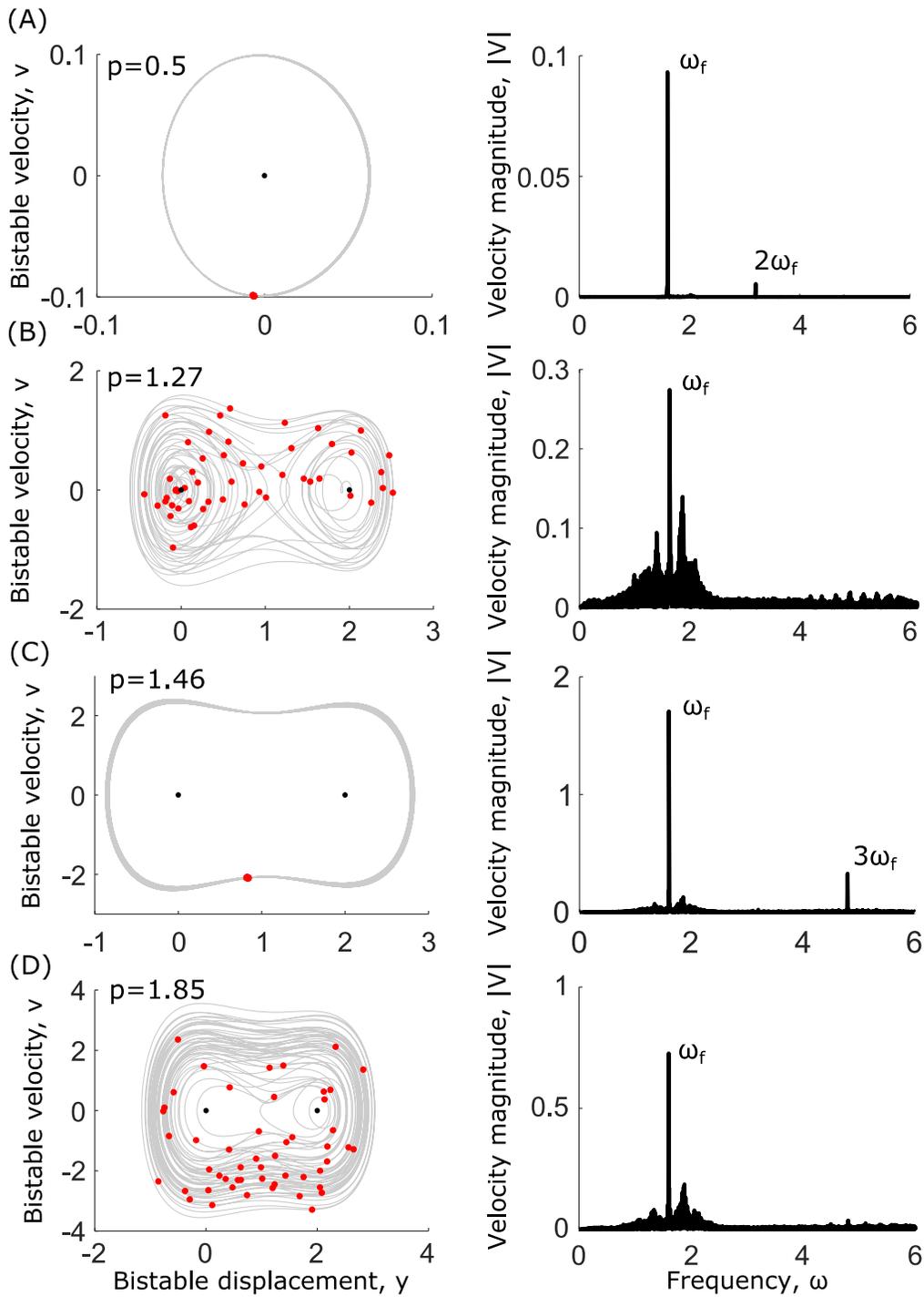

Figure 7. Phase diagram of the bistable constituents of the first unit under forward excitation Figure 2, and corresponding FFT of the bistable velocity for different input excitation levels. (A) through (D) matches the same labels in Figure 6. Black dots represent the equilibrium positions of the bistable constituents and red dots correspond to the stroboscopic map of the first bistable constituent. $\omega = 1.6$ denotes the input frequency.

*3.3 Region III*

Lastly, when excitation frequency is inside region III, we explored the characteristic dynamics with input frequency $\omega = 2.2$, labeled (c) in Figure 3. As indicated by the complicated TR pattern in Figure 8(b), for parameters in region III, we expect the dynamics of the first bistable element to be more intricate. Depicted in Figure 8(a), as input amplitude increases, dynamics of the first bistable element constantly transitions amongst periodic intrawell (point A and F), quasi-periodic intrawell (point B), aperiodic intrawell (point C), chaotic interwell (point D and G) and even subharmonic intrawell oscillation (point E). As a result, transmittance ratio frequently fluctuates up and down and is generally low for periodic intrawell and high for chaotic/aperiodic interwell/intrawell oscillations. In contrast to the previous two cases, for input frequencies in this region, there is no clear threshold on input amplitude beyond which vibration energy always propagates through the chain. However, when energy does propagate through the chain, supratransmission is still enabled via nonlinear instability, exemplified by the sudden change of vibration amplitude of the bistable element.

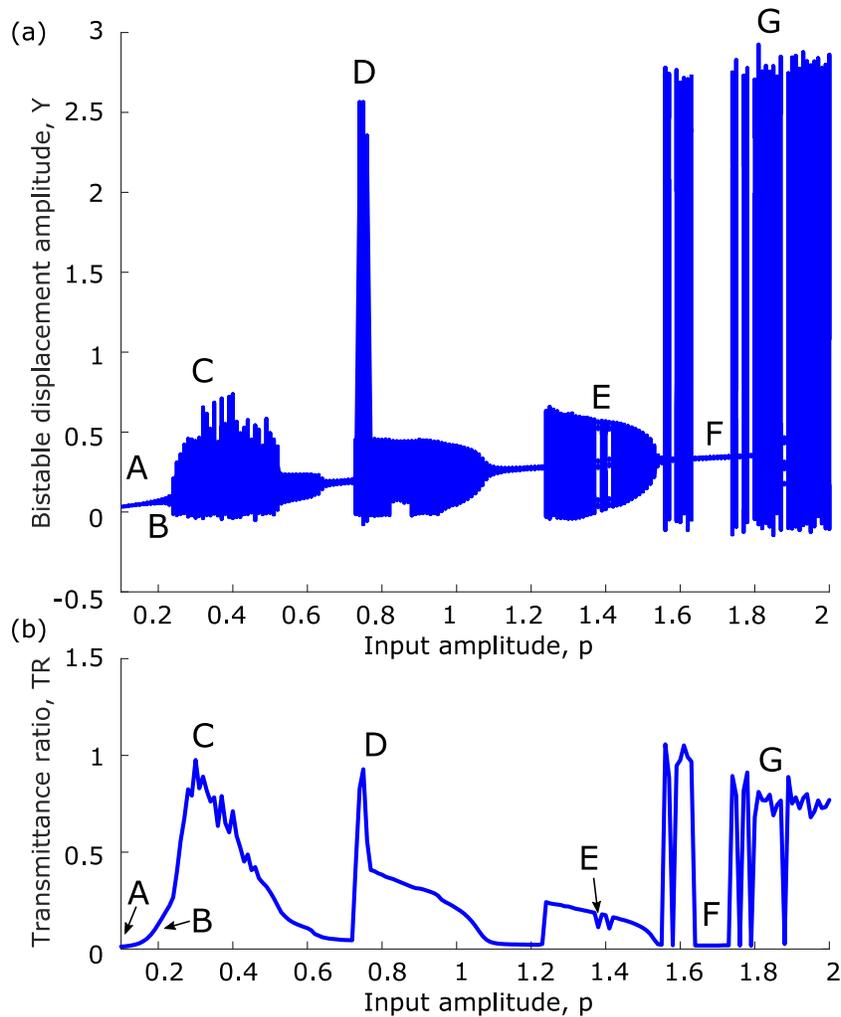

Figure 8. Displacement amplitude of the first bistable constituent and corresponding transmittance ratio as input amplitude increases for forward excitation scenarios with input frequency ω=2.2. Displacement amplitude is determined to be the relative displacement amplitude between the two moving masses at steady state. A through G corresponds to input amplitude $p = [0.1, 0.2, 0.3, 0.75, 1.38, 1.7\ 1.85]$.

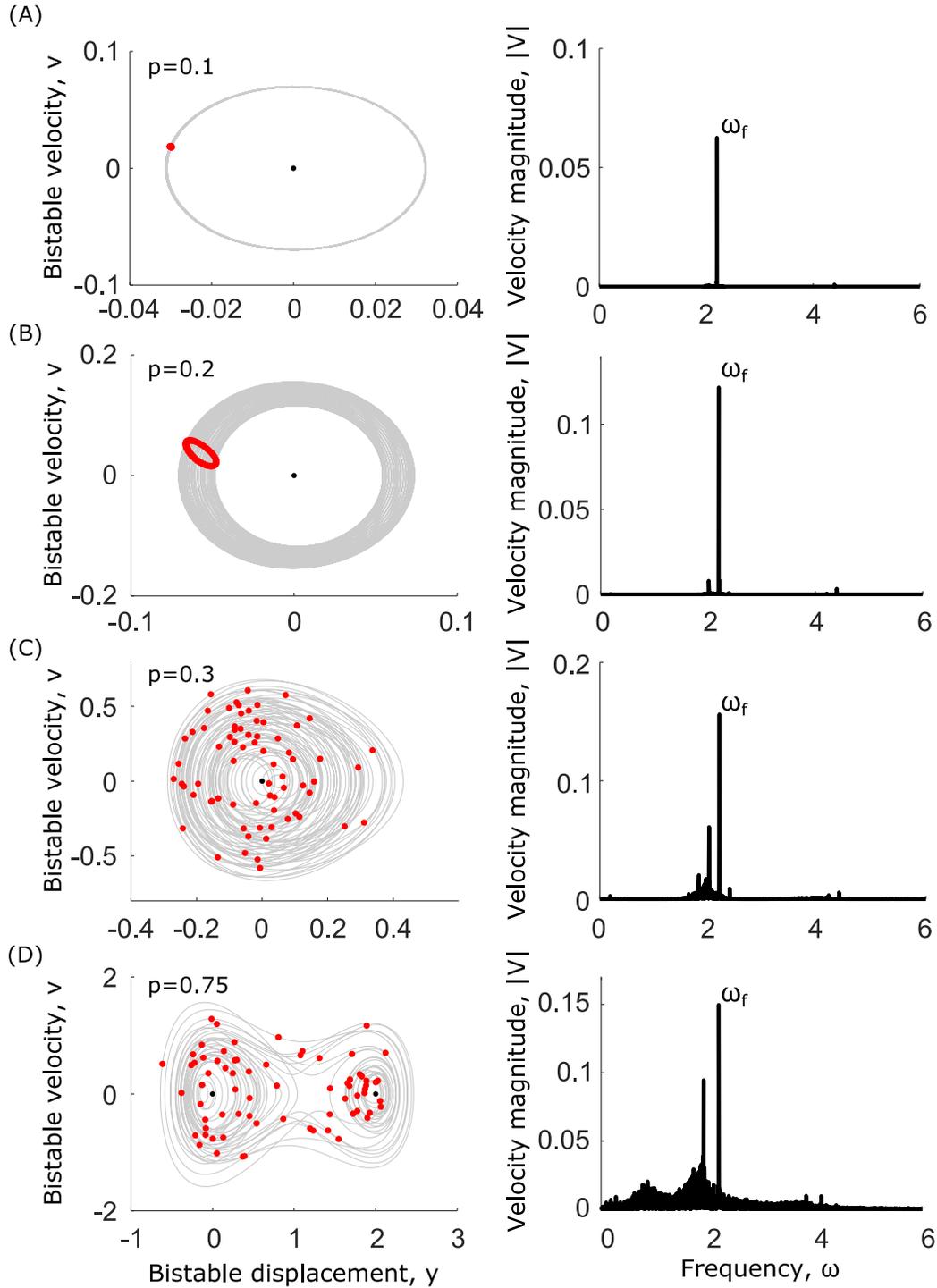

Figure 9. Phase diagram of the bistable constituents of the first unit under forward excitation, Figure 2, and corresponding FFT of the bistable velocity for different input excitation levels. (A) through (G) matches the same labels in Figure 8. Black dots represent the equilibrium positions of the bistable constituents and red dots correspond to the stroboscopic map of the first bistable constituent. $\omega = 2.2$ denotes the input frequency.

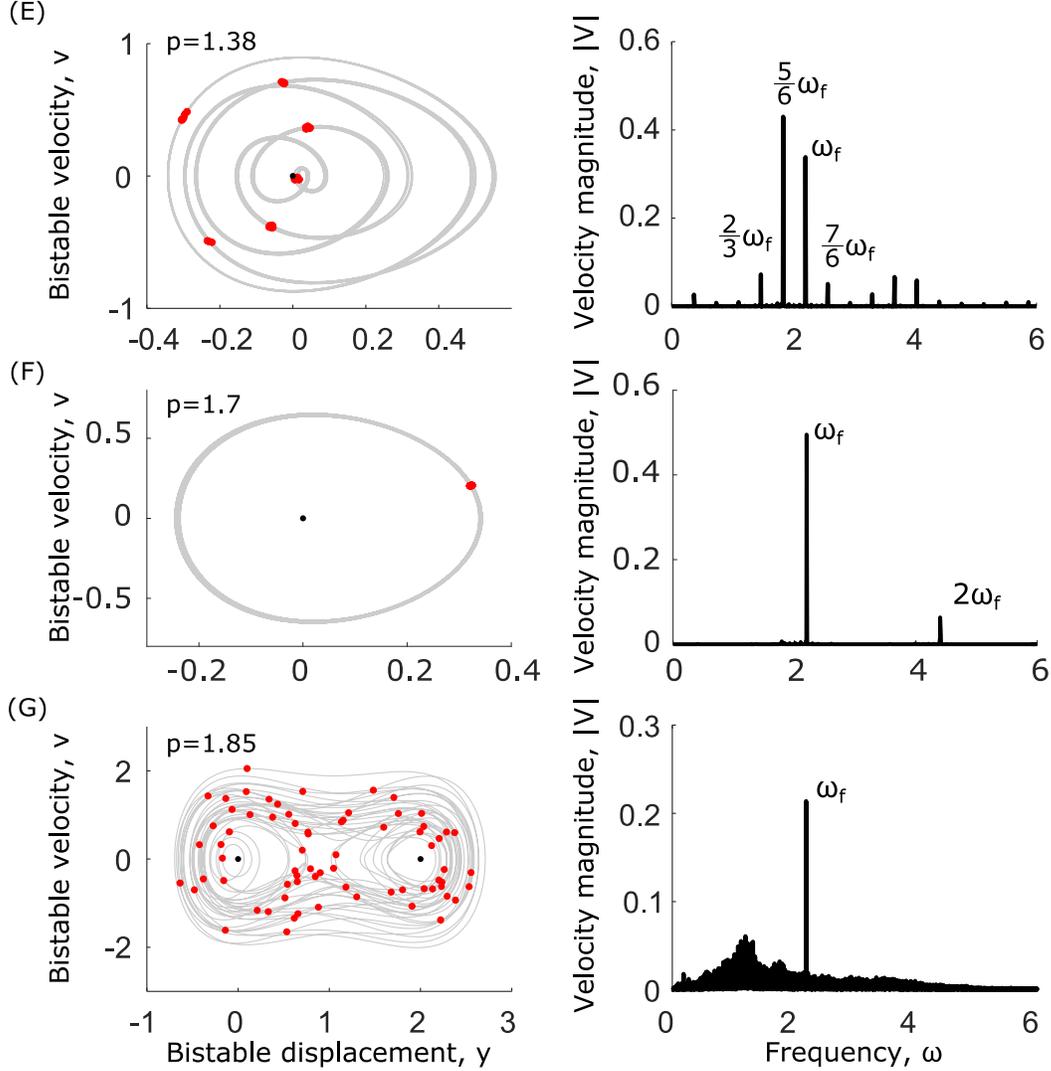

Figure 9. (Cont'd) Phase diagram of the bistable constituent of the first unit under forward excitation Figure 2, and corresponding FFT of the bistable velocity for different input excitation levels.

## 4 Predicting the onset of supratransmission and region of non-reciprocal transmission – An analytical approach

To facilitate the design and synthesis of metastructures with desired non-reciprocal wave energy propagation characteristics, in this study, we seek to develop analytical tools that is able to effectively and efficiently predict the region of non-reciprocal transmission of the proposed metastructure.

The proposed metastructure is a multi-degree of freedom highly nonlinear systems that are known to be challenging and often times impossible to directly conduct analytical study on. On the other hand, from investigations discussed in Sec. 3, we noticeably observed the correlation between dynamics of the first bistable element with respect to the transmission ratio (TR) under forward actuation. This is mainly for two reasons: first is the weak coupling between the individual units (to maintain adaptable and metastable characteristics, strong coupling is not desired [33] [34]) and second is because the driving frequency is inside the stopband, vibration is prone to be localized closer to the excitation initially. Therefore, we propose to predict the onset of supratransmission of the fully nonlinear metsatructure by analyzing an equivalent but simplified mathematical model that only retains the nonlinear element of the first unit while linearizing the rest of the chain. With such a localized nonlinear-linear model, we are endowed with the opportunity to derive analytical solutions that approximate the supratransmission threshold of the fully nonlinear system, by using the transfer matrix method for the linear part [41] and interface with the harmonic balance method of the first nonlinear unit [40]. Similar concepts of treating the nonlinearity locally has been demonstrated to be effective in analyzing vibration characteristics for other nonlinear system [32] [42] [43]. Additionally, we noticed that when the first bistable element transitions from intrawell to interwell oscillation via a saddle node bifurcation (region II), a clear threshold on input amplitude can be identified beyond which drastic increase in transmittance ratio is observed. Such a rapid growth in transmittance ratio could potentially be used as a practical means for identifying regions of non-reciprocal response [21]. Therefore, in the following investigation discussed in this paper, we focus on analytically predicting the onset of supratransmission induced by the saddle-node bifurcation.

*4.1 Forward actuation*

Figure 10(a) depicts the schematic of the localized nonlinear-linear model for forward actuation in which nonlinearity is only preserved in the first unit. Governing equations of the $j^{th}$ unit, Figure 10(b) can be expressed as:

$$m_1\ddot{w}_{[j+1]1} + \tilde{k}_1(w_{[j+1]1} - w_{[j]2}) + k_L(w_{[j+1]1} - w_{[j+1]2}) = 0 \tag{6}$$

$$m_2\ddot{w}_{[j]2} + \tilde{k}_1(w_{[j]2} - w_{[j+1]1}) + k_L(w_{[j]2} - w_{[j]1}) = 0 \tag{7}$$

for $1 \leq j \leq N-1$, where $\tilde{k}_1$ is the linearized stiffness of the bistable constituents and $w_{[j]1}$ and $w_{[j]2}$ are displacements from the new equilibrium positions for a fixed z and configuration.

After nondimensionalization, the governing equations become

$$\mu w''_{[j+1]1} + \tilde{\beta}_1(w_{[j+1]1} - w_{[j]2}) + (w_{[j+1]1} - w_{[j+1]2}) = 0 \tag{8}$$

$$w''_{[j]2} + \tilde{\beta}_1(w_{[j]2} - w_{[j+1]1}) + (w_{[j]2} - w_{[j]1}) = 0 \tag{9}$$

where the parameters are defined as

$$\omega^2 = k_L/m_2 \,;\, \mu = m_1/m_2 \,;\, \tilde{\beta}_1 = \tilde{k}_1/k_L$$

and the operator ( )′ represents a derivative with respect to nondimensional time $t$.

Figure 10(b) shows a schematic representation of two adjacent units in the assembled periodic structure. On both side of each unit, there is a pair of force and displacement. It should be noted that in this case $u_R^j = w_{[j+1]1}$ and $u_R^{j+1} = w_{[j+2]1}$. Balancing force at the left and right ends of the $(j+1)^{th}$ unit, we can derive the following relations:

$$u_R^{(j+1)} = \frac{-\omega^2 + \tilde{\beta}_1}{\tilde{\beta}_1} u_L^{(j+1)} + \frac{\omega^2 - 1 - \tilde{\beta}_1}{\tilde{\beta}_1} u_L^{(j+1)} = T_{11}^F u_L^{(j+1)} + T_{12}^F F_L^{(j+1)}$$

$$F_R^{(j+1)} = \frac{\mu\omega^4 - \omega^2 \tilde{\beta}_1(1+\mu)}{\tilde{\beta}_1} u_L^{(j+1)} + \frac{-\mu\omega^4 + \omega^2(\tilde{\beta}_1 + \mu + \tilde{\beta}_1\mu) - \tilde{\beta}_1}{\tilde{\beta}_1} F_L^{(j+1)} = T_{21}^F u_L^{(j+1)} + T_{22}^F F_L^{(j+1)}$$

From compatibility and force equilibrium between adjacent units we have $u_L^{(j+1)} = u_R^{(j)}$ and $F_L^{(i+1)} = -F_R^{(j)}$. The transfer matrix from $j^{th}$ unit to $(j+1)^{th}$ unit can then be derived as:

$$\begin{bmatrix} u_R^{(j+1)} \\ F_R^{(j+1)} \end{bmatrix} = \begin{bmatrix} T_{11}^F & -T_{12}^F \\ T_{21}^F & -T_{22}^F \end{bmatrix} \begin{bmatrix} u_R^{(j)} \\ F_R^{(j)} \end{bmatrix}$$

Moving along the finite chain from the second to last unit to the first unit, we can write

$$\begin{bmatrix} u_R^{(N-2)} \\ F_R^{(N-2)} \end{bmatrix} = \begin{bmatrix} T_{11}^F & -T_{12}^F \\ T_{21}^F & -T_{22}^F \end{bmatrix}^{N-2} \begin{bmatrix} u_R^{(1)} \\ F_R^{(1)} \end{bmatrix} = \begin{bmatrix} M_{11}^F & M_{12}^F \\ M_{21}^F & M_{22}^F \end{bmatrix} \begin{bmatrix} u_R^{(1)} \\ F_R^{(1)} \end{bmatrix} \tag{10}$$

For last unit, Figure 10(c), with fixed boundary conditions, governing equations can be expressed as:

$$\mu w''_{[N]1} + \tilde{\beta}_1(w_{[N]1} - w_{[N-1]2}) + (w_{[N]1}) = 0 \tag{11}$$

$$w''_{[N-1]2} + (w_{[N-1]2} - w_{[N-1]1}) + \tilde{\beta}_1(w_{[N-1]2} - w_{[N]1}) = 0 \tag{12}$$

It is then easy to get that $F_L^{(N-1)} = C_1 u_L^{(N-1)}$ where $C_1 = \frac{(\omega^4 \mu - \omega^2(1+\tilde{\beta}_1+\tilde{\beta}_1\mu)+\tilde{\beta}_1)}{\omega^4\mu - \omega^2(1+\tilde{\beta}_1)(1+\mu)+(2+\tilde{\beta}_1)}$. Combining with Eq. (10), we can derive that

$$F_R^{(1)} = Cu_R^{(1)} = \frac{-C_1 M_{11}^F - M_{21}^F}{(C_1 M_{12}^F + M_{22}^F)} u_R^{(1)} \tag{13}$$

Lastly, governing equations for the first nonlinear unit, Figure 10(d) can be derived as:

$$\mu(x + y + x_{in})'' + \beta_1 y + \beta_2 y^2 + \beta_3 y^3 = F_R^1 \tag{14}$$

$$(x + x_{in})'' + (x) - (\beta_1 y + \beta_2 y^2 + \beta_3 y^3) = 0 \tag{15}$$

where $x$ and $y$ are deformations of linear and nonlinear springs and $\beta_1 = k_1/k_L$, $\beta_2 = k_2/k_L$, and $\beta_3 = k_3/k_L$. Without loss of generality, we assume $\beta_1 = 2\beta_3$ and $\beta_2 = -3\beta_3$ to realize a symmetric bistable potential with distance between the two equilibria to be 2, similar to the potential profile used in previous numerical analysis.

Assuming 1-term expansion with slowly varying coefficients $x(t) = a(t)\sin(\omega t) + b(t)\cos(\omega t)$ and $y(t) = c(t)\sin(\omega t) + d(t)\cos(\omega t) + e(t)$, plugging equation (13), $u_R^1 = x_{in} + x(t) + y(t)$ and $x_{in} = p\sin(\omega t)$ into the equations of motion, eliminating higher order terms, and grouping the constant, $\sin(\omega t)$ and $\cos(\omega t)$ terms yields five equations for the coefficients, $a(t)$, $b(t)$, $c(t)$, $d(t)$ and $e(t)$

$$2\omega b' = -p\omega^2 + (1-\omega^2)a(t) + \left(-2 - \tfrac{3}{4}R_y^2 + 6e(t) - 3e(t)^2\right)\beta_3 c(t)$$

$$-2\omega a' = (1-\omega^2)b(t) + \left(-2 - \tfrac{3}{4}R_y^2 + 6e(t) - 3e(t)^2\right)\beta_3 d(t)$$

$$2\mu\omega b' + 2\mu\omega e' = (-C - \mu\omega^2)p + (-C - \mu\omega^2)a(t) + \left(2\beta_3 + \tfrac{3}{4}R_y^2 - C - \mu\omega^2 - 6\beta_3 e(t) + 3\beta_3 e(t)^2\right)c(t)$$

$$-2\mu\omega c' - 2\mu\omega a' = (-C - \mu\omega^2)b(t) + \left(2\beta_3 + \tfrac{3}{4}R_y^2 - C - \mu\omega^2 - 6\beta_3 e(t) + 3\beta_3 e(t)^2\right)d(t)$$

$$0 = -2\beta_3 e(t) + 3\beta_3 e(t)^2 - \beta_3 e(t)^3 + \left(\tfrac{3}{2} - \tfrac{3}{2}e(t)\right)\beta_3 R_y^2$$

where $R_y$ is the bistable displacement amplitude with $R_y^2 = c(t)^2 + d(t)^2$

Combining the five equations, a third order polynomial in terms of $R = R_y^2$ can be derived [40]:

$$\rho_3 R^3 + \rho_2 R^2 + \rho R + \rho_0 = 0$$

where

$$\rho_3 = \tfrac{1}{16(-1+\omega^2)^2}\left(225\,C^2\beta_3^2 + 450\,C\beta_3^2(-1+(1+\mu)\omega^2) + 225\beta_3^2(-1+(1+\mu)\omega^2)^2\right)$$

$$\rho_2 = \tfrac{1}{16(-1+\omega^2)^2}\big(-120\,C^2\beta_3 - 240\,C^2\beta_3^2 + 120\,C^2\beta_3\omega^2 - 120\beta_3(-1 + (1+\mu)\omega^2)(-\mu\omega^2(-1+\omega^2) + 2\beta_3(-1+(1+\mu)\omega^2)) - 120\,C\beta_3(4\beta_3(-1+(1+\mu)\omega^2) - (-1+\omega^2)(-1+(1+2\mu)\omega^2))\big)$$

$$\rho_1 = \tfrac{1}{16(-1+\omega^2)^2}\big(64\,C^2 - 128\,C^2\omega^2 + 64\,C^2\omega^4 - 128\,C^2\beta_3 + 128\,C^2\omega^2\beta_3 + 64C^2\beta_3^2 + 16\left(\omega^2(-1+\omega^2)\mu - 2\beta_3(-1+\omega^2(1+\mu))\right)^2 + 64\,C(-1+\omega^2+\beta_3)(-\omega^2(-1+\omega^2)\mu + 2\beta_3(-1+\omega^2(1+\mu)))\big)$$

$$\rho_0 = \tfrac{1}{16(-1+\omega^2)^2}(-16C^2 p^2 - 32Cp^2\mu\omega^2 - 16p^2\mu^2\omega^4)$$

This polynomial can be further simplified as $R^3 + a_2 R^2 + a_1 R + a_0 = 0$ with leading coefficients to be 1. Transmission threshold corresponding to a saddle-node bifurcation can then be determined by solving the equation

$$a_0 = -2 \times (P + \sqrt{Q^3}) \tag{16}$$

where $P = (a_2^3)/27 - (a_1 a_2)/6$ and $Q = (a_2^2)/9 - a_1/3$

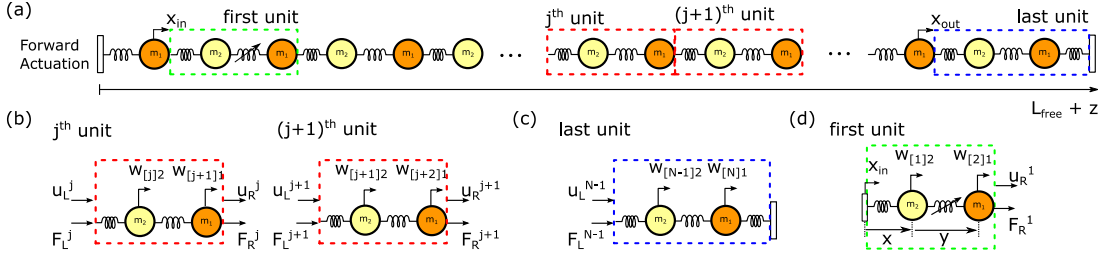

Figure 10.(a) Discrete chain of localized nonlinear-linear model for forward actuation; Only the first unit adjacent to input is nonlinear (b) Schematic of $j^{th}$ and $(j+1)^{th}$ unit for transfer matrix analysis. Subscript in square bracket corresponds to the index of metastable module shown in Figure 2. (c) Schematic of the last unit for transfer matrix analysis. (d) Schematic of first unit for harmonic balancing analysis.

## 4.2 Backward actuation

Following the same procedure, transfer matrix from the second to last unit to the first unit for backward actuation, Figure 11, can be derived as

$$\begin{bmatrix} u_R^{(N-2)} \\ F_R^{(N-2)} \end{bmatrix} = \begin{bmatrix} T_{11}^B & -T_{12}^B \\ T_{21}^B & -T_{22}^B \end{bmatrix}^{N-2} \begin{bmatrix} u_R^{(1)} \\ F_R^{(1)} \end{bmatrix} = \begin{bmatrix} M_{11}^B & M_{12}^B \\ M_{21}^B & M_{22}^B \end{bmatrix} \begin{bmatrix} u_R^{(1)} \\ F_R^{(1)} \end{bmatrix} \tag{17}$$

where $T_{11}^B = -\omega^2 + 1$, $T_{12}^B = (\omega^2 - 1 - \tilde{\beta}_1)/\tilde{\beta}_1$, $T_{21}^B = \omega^4 \mu - \omega^2(1+\mu)$, and $T_{22}^B = (-\omega^4 \mu + \omega^2(1+\mu+\tilde{\beta}_1 \mu) + \tilde{\beta}_1)/\tilde{\beta}_1$. With fixed boundary conditions, we can derive that

$$F_R^{(1)} = C u_R^{(1)} = \frac{-C_1 M_{11}^F - M_{21}^F}{(C_1 M_{12}^F + M_{22}^F)} u_R^{(1)} \tag{18}$$

where $C_1 = \frac{\tilde{\beta}_1 (\tilde{\beta}_1 + \omega^4 \mu - \omega^2(1+\tilde{\beta}_1+\mu))}{(\tilde{\beta}_1 (2+\tilde{\beta}_1) + \omega^4 \mu - \omega^2(1+\tilde{\beta}_1)(1+\mu))}$.

For the first nonlinear unit, Figure 11(d), assuming 1-term expansion with slowly varying coefficients $x(t) = a(t)\sin(\omega t) + b(t)\cos(\omega t)$ and $y(t) = c(t)\sin(\omega t) + d(t)\cos(\omega t) + e(t)$, plugging equation (17), $u_R^1 = x_{in} + x(t) + y(t)$ and $x_{in} = p\sin(\omega t)$ into equations of motion, eliminating higher order terms, grouping the constant, $\sin(\omega t)$ and $\cos(\omega t)$ terms and combining the equations, a third order polynomial can be similarly derived for $R = R_y^2$ where $R_y$ is the bistable displacement amplitude with $R_y^2 = c(t)^2 + d(t)^2$:

$$\rho_3 R^3 + \rho_2 R^2 + \rho R + \rho_0 = 0$$

where

$$\rho_3 = \frac{225}{16} \beta_3^2$$

$$\rho_2 = \frac{1}{16(-1+C+\omega^2\mu)^2} (120\,\beta\,(-1+C+\omega^2\mu)\,(C(-1+\omega^2-2\beta)+2\beta+\omega^4\mu-\omega^2(1+\mu+2\beta\mu)))$$

$$\rho_1 = \frac{1}{16\,(-1+C+\omega^2\mu)^2}\,\big(16\,\big(C(-1+\omega^2-2\beta)+2\beta+\omega^4\mu-\omega^2(1+\mu+2\beta\mu)\big)^2\big)$$

$$\rho_0 = \frac{1}{16\,(-1+C+\omega^2\mu)^2}\,\big(-16\,p^2\,\big(C(-1+\omega^2)+\omega^2(-1+(-1+\omega^2)\mu)\big)^2\big)$$

This polynomial can be further simplified as $R^3 + a_2 R^2 + a_1 R + a_0 = 0$ with leading coefficients to be 1. Transmission threshold of the bifurcation can then be determined by solving the same Eq.(16) with newly derived coefficients $a_1, a_2$ and $a_3$ for backward actuation.

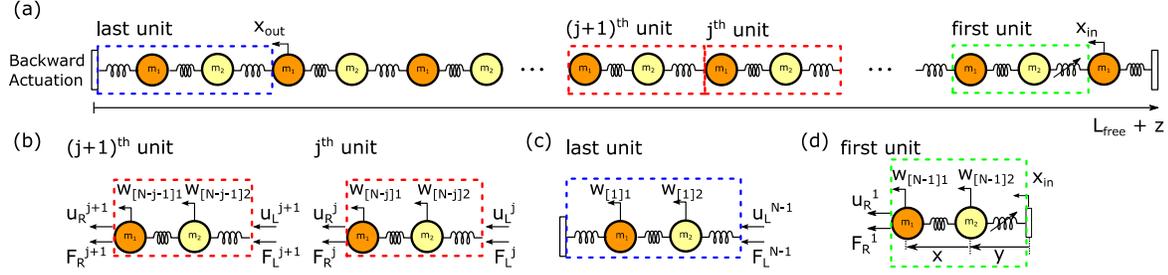

Figure 11.(a) Discrete chain of localized nonlinear-linear model for backward actuation; Only the first unit adjacent to input is nonlinear (b) Schematic of $j^{th}$ and $(j+1)^{th}$ unit for transfer matrix analysis. Subscript in square bracket corresponds to the index of metastable module shown in Figure 2. (c) Schematic of the last unit for transfer matrix analysis. (d) Schematic of first unit for harmonic balancing analysis.

*4.3 Analytical results*

With the localized nonlinear-linear method described in Sec 4.1 and 4.2 for forward and backward actuations, we compare the analytical prediction of the simplified model with numerical investigations of the fully nonlinear metastructure to evaluate the effectiveness of the analytical method. Parameters used for simulations are the same as in Sec. 3. Figure 12 depicts the contour plot on transmittance ratio as input frequency and amplitude varies for both forward and backward actuations, with dark area corresponds to non-propagation zone. Therefore, the boundaries of supratransmission in the parameter space correspond to the transitions of color from dark to bright. Solid red and black lines are analytical predicted onset of supratransmission for forward and backward actuations, respectively. As shown in Figure 11(a), the proposed method is able to predict accurately the onset of the supratransmission induced by the saddle-node bifurcation in region II discussed in Sec. 3 for forward actuation over a range of the parameters investigated. For backward actuation, analytical predicted boundary (black lines) also closely resembles the trend of the numerical simulation: input amplitude level required to facilitate the onset of supratransmission increases as input frequency decreases. Combining the two predicted boundaries, Figure 12(b), since it predicts that wave energy can only propagate through the chain with input parameters above the red (black) line for forward (backward) actuation scenario, there exists a region in parameter space that are bounded by the two predicted transition boundaries such that wave energy can only transmit in one direction. Hence, the proposed metastable structure is capable of attaining supratransmission at different excitation level depending on the excitation directions. Therefore, by combining the supratransmission property of a nonlinear periodic chain with spatial asymmetry, we are able to enable non-reciprocal wave propagation and predict the non-reciprocal region using the proposed analytical approach. Overall, the localized nonlinear-linear model and prediction method proposed in the study are beneficial for identifying the unidirectional wave propagation region in the input parameter space, which can greatly reduce the computational time and complexity.

It should be noted that in this study, as a proof of concept, we have focused on analytically predicting the onset of supratransmission induced mainly by the saddle-node bifurcation for one of the metastable states. On the other hand, as discussed in Sec. 3, onset of supratransmission can be facilitated through other intricate nonlinear instabilities as well, which could be predicted using similar approach. That is, utilizing the same principle of localized nonlinear-linear model, we could adopt different forms of harmonic expansion for the first nonlinear unit to derive other supratransmission boundaries of interest [44] [45] [46] [47]. Additionally, the proposed analytical method can be readily extended to other metastable states by linearizing the system around different starting positions.

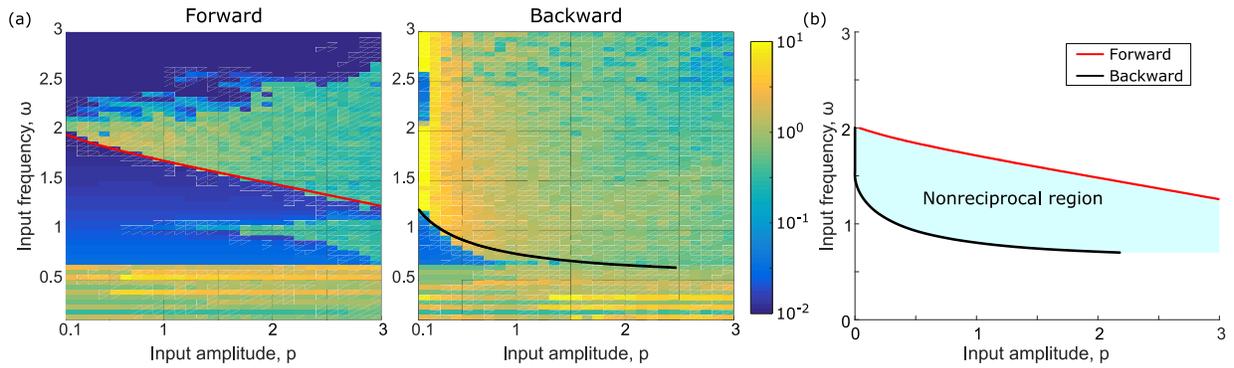

Figure 12. (a) Contour plot on transmittance ratio (vs. input frequency and amplitude) of forward and backward actuations for free length configuration with analytically predicted supratransmission boundaries. (b) Non-reciprocal region is bounded in between the two analytically predicted boundaries.

## 5  Parametric studies on supratransmission threshold

From both numerical and analytical investigations in Sec. 3 and 4, we find that when the excitation frequency is within the stopband of the metastructure, there exhibits a threshold input amplitude beyond which wave energy is able to propagate through the chain due to supratransmission. Additionally, when this onset of supratransmission is induced by saddle-node bifurcation, a drastic increase in transmittance ratio can be observed as input amplitude exceeds the threshold value and this sudden change could potentially be used as a practical means to design non-reciprocal regions. Therefore, in the following sections, we seek to explore influences of important parameters on the threshold amplitude to trigger supratransmission and to provide more insight for synthesizing systems with desired non-reciprocal characteristics.

*5.1 Influence of damping coefficient*

In this section, the influence of damping coefficients $c$ on the threshold amplitude is studied. System parameters used are the same as in Sec 3 and 4 and excitation frequency is chosen to be $\omega = 1.6$. For exploration purposes, the metastructure is actuated under forward excitation scenario with damping coefficients $c$ varying from 0.001, 0.005, 0.01, 0.05 to 0.1.

Figure 13 depicts the transmittance ratio as input amplitude increases for different damping coefficients. With small damping coefficients, $c = 0.001$, $c = 0.005$, and $c = 0.01$, threshold amplitudes to trigger the onset of supratransmission gradually increase from $p = 1.27$ to $p = 1.28$ and $p = 1.31$. Additionally, due to the increased damping coefficient, energy transmitted through the structure in general decreases hence the transmittance ratio also decreases. As the damping ratio further increases to $c = 0.05$ and $c = 0.1$, the threshold amplitude required to enable supratransmission increases to 1.54 and 1.75 respectively and the transmitted energy drops drastically with transmittance ratio (TR) decreases from more than 1 to approximately 0.26. Therefore, increasing damping coefficients will increase the threshold amplitude required to trigger the onset of supratransmission and decrease the amount of energy propagated through the structure.

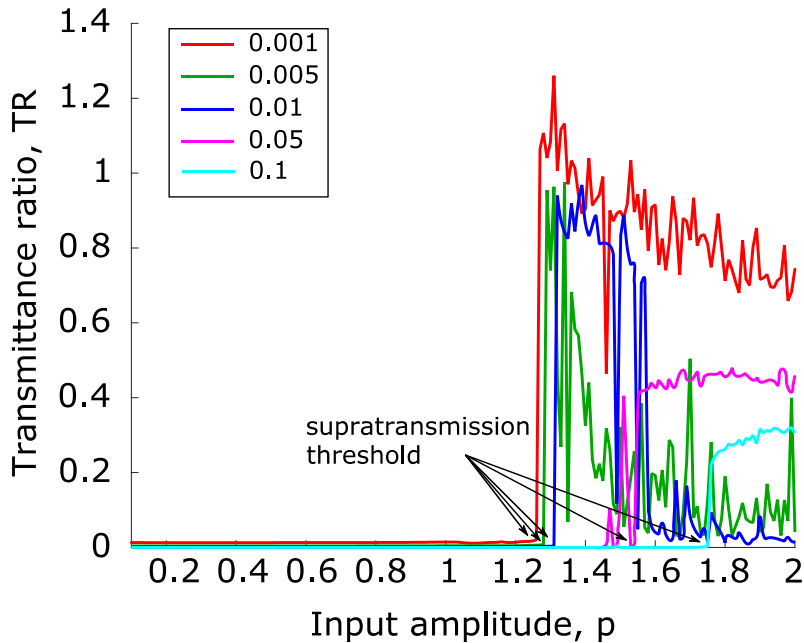

Figure 13. Transmittance ratio of metastructure under forward excitation as input amplitude increases for different damping coefficients $c = [0.001, 0.005, 0.01, 0.05, 0.1]$.

## 5.2 Influence of the number of modules

Results presented in the previous sections are all based on a 10 module assembly, in this section, the influence of module number $N$ on the threshold amplitude is studied. System parameters used are the same as those used in Sec 3 and 4 with damping coefficients $c = 0.001$. Again, for exploration purposes, the metastructue is actuated only under the forward excitation scenario.

Figure 14 depicts the threshold amplitude that enables supratransmission as module number increases from 5, 10, 20, 50 to 100 for a range of frequencies inside Region II, Figure 3. For a fixed excitation frequency, threshold amplitude initially may fluctuates with small module number and as the number of modules increases, this value becomes constant and is invariant of module number. Such an invariant characteristic is similarly observed in other nonlinear metamaterials [48]. Additionally, from the study of the 10 module assembly, we noticed that as the input frequency increases, the threshold amplitude to trigger supratransmission deceases, Figure 12. This trend is also preserved as module number changes from 5 to 100: the threshold amplitude requires to facilitate supratransmission always decreases as frequency increases.

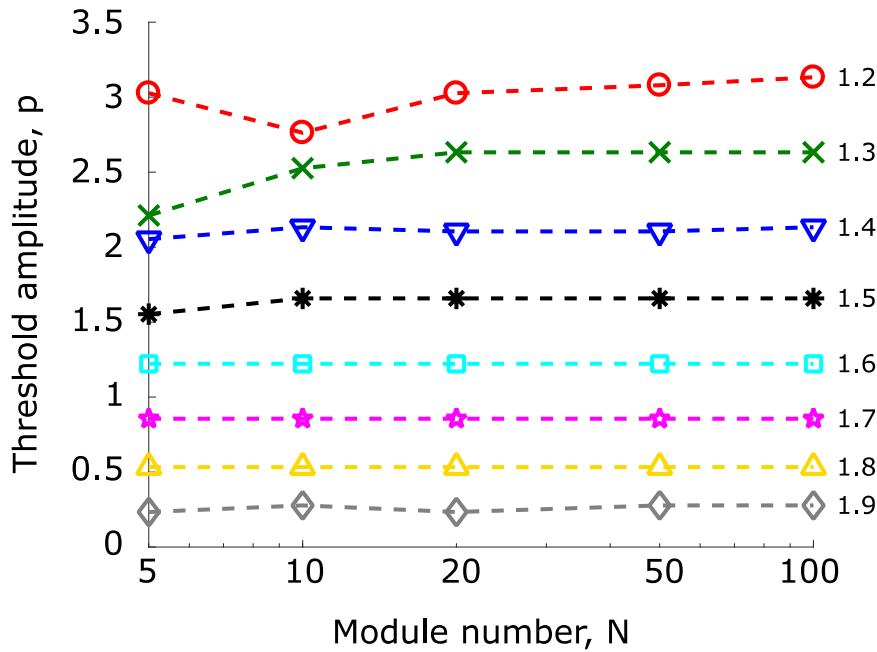

Figure 14. Threshold amplitude of metastructure under forward excitation for module number $N = [5, 10, 20, 50, 100]$ and excitation frequency $\omega = [1.2, 1.3, 1.4, 1.5, 1.6, 1.7, 1.8, 1.9]$.

## 6   Conclusion

In this paper, we numerically and analytically explore the supratransmission phenomenon in a metastable modular metastructure. From numerical investigations on a 1D metastable modular chain, we explore intricate dynamics affordable by the metastructure and elucidate different kinds of nonlinear instabilities that facilitate the onset of supratransmission. In the case of saddle-node bifurcation instability, the transmission ratio could suddenly increase significantly as the input amplitude increases and exceeds certain threshold. We find that when the excitation frequency is inside the stopband, dynamics of the first nonlinear constituent close to the input can considerably influence the wave energy transmission characteristics of the overall system. Therefore, we propose a localized nonlinear-linear model to analytically estimate the input threshold amplitude required to trigger supratransmission of the nonlinear metastable modular chain. Through intelligently integrating the supratransmission property in a nonlinear periodic chain with spatial asymmetry, the metastable structure is capable of realizing non-reciprocal wave propagation by attaining sudden propagation of wave energy at different excitation levels depending on the input directions. Utilizing the proposed analytical tool, we are able to accurately and efficiently predict the non-reciprocal region depending on input frequency and amplitude. Additionally, we explore the influence of system damping and module number on the threshold amplitude to trigger the onset of supratransmission. The outcomes of this study provide effective methodology and great potential to design systems with desired adaptable non-reciprocal wave energy propagation characteristics utilizing the proposed reconfigurable metastable modular metastructure concept.

**Acknowledgment**

The authors gratefully acknowledge the support of the U.S. Army Research Office under grant number W911NF-15-1-0114.